# OPEN DROPLET MICROFLUIDICS FOR TESTING MULTI-DRUG RESISTANCE AND ANTIBIOTIC RESILIENCE IN BACTERIA

*Santosh Pandey[1], Taejoon Kong[1], Nicholas Backes[2] and Gregory J. Phillips[2]*
[1]Electrical and Computer Engineering, Iowa State University, Ames, Iowa, USA
[2]Veterinary Microbiology and Preventive Medicine, Iowa State University, Ames, Iowa, USA

## ABSTRACT

New combinations of existing antibiotics are being investigated to combat bacterial resilence. This requires detection technologies with reasonable cost, accuracy, resolution, and throughput. Here, we present a multi-drug screening platform for bacterial cultures by combining droplet microfluidics, search algorithms, and imaging with a wide field of view. We remotely alter the chemical microenvironment around cells and test 12 combinations of resistant cell types and chemicals. Fluorescence intensity readouts allow us to infer bacterial resistance to specific antibiotics within 8 hours. The platform has potential to detect and identify parameters of bacterial resilience in cell cultures, biofilms, and microbial aggregates.

## INTRODUCTION

### Antimicrobial Resistance and Detection Strategies

Antimicrobial resistance is an alarming threat to global public health [1]. There are a limited number of drugs approved by the World Health Organization. The time and cost for new drug discovery are significant, and existing therapy options are becoming limited [2]. In addition to the challenge of drug resistance, the inherent heterogeneity of bacterial cells at the population level (e.g. in cell growth, cell division, or gene expression) gives rise to the phenomena of resilience where bacteria can survive antibiotic exposure by entering into a non-growth state [1]. This makes it difficult to develop new strategies to completely eliminate bacterial pathogens. The delay in detection and control of bacterial pathogens, for example in livestock farming, often leads to infection in farm animals that can sicken the entire herd and result in extra medical costs [3]. Bacterial pathogens in meat-producing animals can make their way through the food supply chain into humans who consume under-cooked meat or poultry products. Large-scale treatment efforts on infected people is non-trivial as the nature and degree of bacterial infection is multi-dimensional and convoluted.

To address the health concerns of antimicrobial resistance, researchers are pursuing novel strategies in detection technologies, such as the development of sensitive chemical probes, new devices for monitoring microbial growth, imaging single cells, instrumentation, robotics, optics, and data analytics [4]. These methods are not without their limitations, however. For example, while culture-based assays, such as agarose plate assays, disc diffusion and broth dilution tests, can have high throughput to test drug resistance levels, they can be relatively slow to yield results. Microfluidic platforms allow the access and control of the chemical microenvironment around cells [5], but have limited flexibility in testing a broad range of chemical combinations for extended times (greater than 6 hours). Automated pipetting robots are expensive and not suitable for wide adoption.

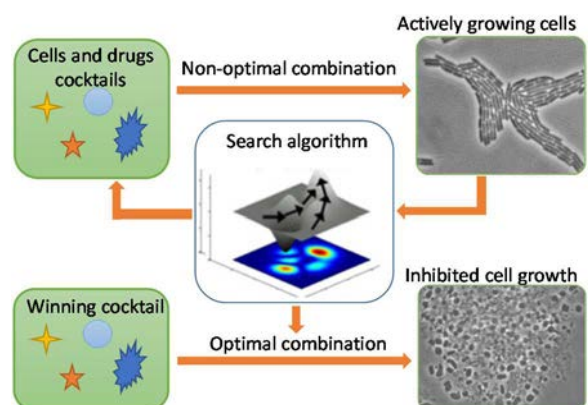

*Figure 1: Overview of the project motivation. The effect of drug cocktails is tested on different bacterial populations. The cell survival percentage is fed into a search algorithm. By using a directed evolutionary search through successive iterations, the winning cocktail (i.e., the one that causes cell death using least concentration of antibiotics) is found.*

There is a need for novel bacterial assays that reduce the time to detect bacterial resilience, require small volumes of resources, and offer significant experimental flexibility to test multiple combinations of antibiotics and healthy or resilient cell populations (Figure 1 and Figure 2) [6-10]. Open microfluidic systems has garnered interest for drug screening where discrete droplets are generated and processed [6]. The methods to produce and transport discrete droplets on open microfluidic systems can either use electric fields to modify the wettability of droplets or employ other physical forces (e.g. magnetic, acoustic, pneumatic or gravitational forces) to control the droplets.

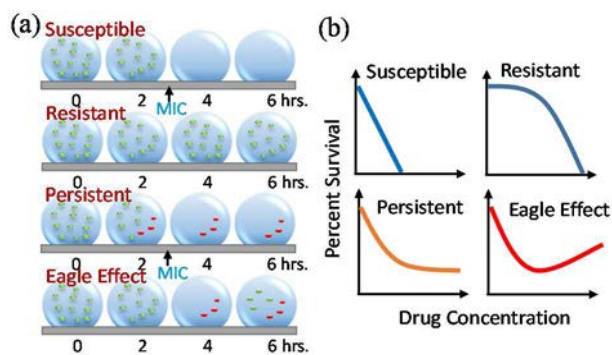

*Figure 2: (a) Types of bacterial resilience. (b) Hypothetical plots of percentage cell survival versus drug concentration for the different types of bacterial resilience.*

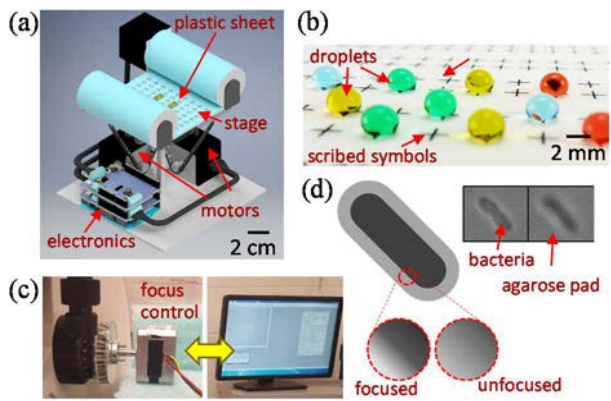

*Figure 3: (a) Structural components of our droplet microfluidic system. (b) Image of the plastic sheet with scribed hydrophilic symbols where droplets are automatically positioned against a superhydrophobic surface. (c-d) Autofocussing is controlled by a GUI software to remotely adjust the microscope's focus to view the cells clearly over long time period (up to 7-10 hours).*

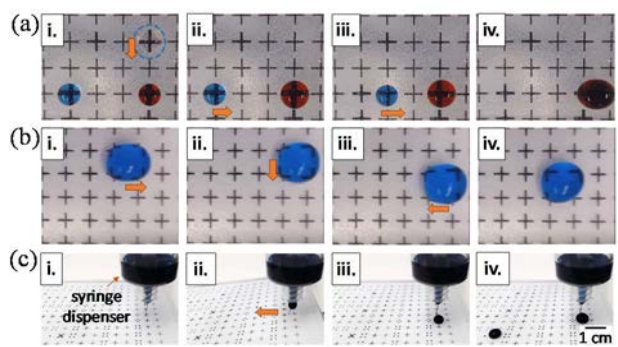

*Figure 4: (a) Merging of three droplets is done by mechanical agitation (orange arrows). The colorless droplet merges with the stationary red droplet, followed by the blue droplet. (b) Large droplet transport is done by using 2 × 2 arrays of scribed symbols. (c) A syringe dispenser setup produces a series of fixed-volume droplets.*

Existing electro-wetting methods require high voltages (100-200 volts) that may produce unknown effects on the biomolecules or cells housed inside droplets. As every electrode is electrically addressable, there is only a finite number of droplets that can be manipulated in parallel. On the other hand, non-electrical methods for droplet manipulation are premature and there may be issues incorporating magnets, optics, sound transducers, or pressure actuators for guiding a large number of droplets.

### Scope of this work

We demonstrate an open microfluidic technology, involving motorized actuation of discrete droplets, as a versatile bacterial resistance screening assay (Figure 3). Various droplet operations are remotely combined to modulate the chemical microenvironments of bacterial populations, along with real-time imaging of their antibiotic resilience. Several chemical microenvironments are tested in one experiment by using automated and remote-controlled handling of droplet operations. An optimization search algorithm is shown to speed up the selection of superior chemical combinations that reduce the cell survival percentage. Without the algorithm-guided search, a very large number of combinations would have to be tested which is impractical on plate assays [9-11]. A scanner-based imaging system is constructed to validate the cell growth in plate assays. In short, our work here combines the technologies of liquid droplet handling, gel gradient microfluidics, CRISPR gene editing, and algorithmic-guided drug combination search for a new application to multi-drug resistance testing.

## MATERIALS AND METHODS
### Droplet Microfluidic System

The droplet microfluidic system consists of three PlexiGlass structural components: base, column, and top stage with plastic sheet. The base is physically screwed to the column. A universal joint connects the column to the top stage, allowing two-axis stage rotation about a central pivot. The system dimensions are: upper stage (9 cm × 9 cm × 1.3 cm), vertical column (20 cm × 20 cm × 0.5 cm), and base (20 cm × 20 cm × 0.5 cm). We have the ability to scale down the dimensions of each part as the system is custom-built in our laboratory [5]. Plastic sheets are coated with a superhydrophobic layer (i.e. Rust-Oleum NeverWet) and scribed with periodic symbols (i.e. +, >, <) using a laser cutter. A roll of plastic sheet may be operated by a stepper motor to dispense fresh sheets before the start of an experiment. Initially, a section of the plastic sheet is taped on to the top stage. The mechanical agitation of the top stage is performed by two stepper motors, where each motor controls one axis of rotation of the top stage. An Arduino microcontroller controls the timing and direction of both stepper motors (NEMA-17, 12 volts, 350 mA, 200 steps per revolution, bipolar mode). A graphical user interface (GUI) provides remote access to the system. Commands to tilt the stage up/down, left/right, and any sequence of such commands are entered in the GUI and transmitted to the microcontroller through its USB port.

### Droplet Dispensing, Transport, and Mixing

The movement of individual droplets is accomplished by mechanical agitation and gravity. The droplets are confined to the scribed symbols on the plastic sheet until the top stage is agitated (or rapidly tilted) by a pre-specified angle. Thereafter, the droplets are able to 'hop' to neighboring scribed symbols by gravity. This rapid tilting action is uniquely suited for droplet transport where the stage is tilted clockwise (or anticlockwise) by a certain angle followed by reverse-tilting the stage anti-clockwise (or clockwise) to its horizontal position. Rapid acceleration and deceleration of a droplet is accomplished by small tilting angles (3°–5°). By varying the shape, size, and location of scribed symbols on the plastic sheet, we can control the direction and movement of individual droplets. Various droplet operations are demonstrated in Figure 4, such as transport, mixing, merging, and dispensing of chemicals to bacteria populations. The actuation system handles operations of over 150 discrete droplets in a single run. There is no need of very high voltages or advanced microfabrication tools integral to digital microfluidics.

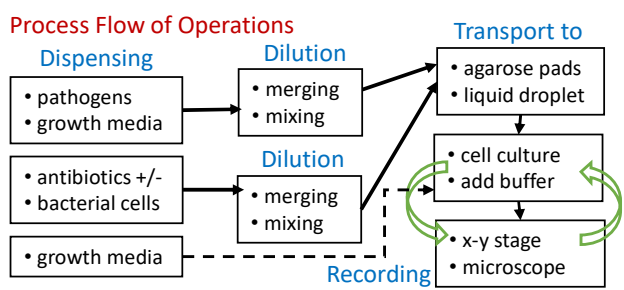

### Imaging with Wide Field-of-View

An upright microscope with a high-speed digital camera (QIImaging QiCam Fast 1394) was used to image the agarose gel droplets or liquid droplets for turbidity. We have a motorized x-y stage for the microscope to accurately position a specific droplet underneath the lens of the microscope. The programmable feature of the x-y stage allows us to record images and videos of a series of agarose pads or liquid droplets in a repeated manner.

We built an autofocusing module to self-correct for any movements in the focal plane that can blur the image. The shift in focus often occurs because of the weight of the microscope's lens and upper body. The autofocusing module instructs the camera to capture ten images at different focal planes. The focus-measure values are calculated for each focal plane. Thereafter, the focus knob is rotated to the focal plane having the maximum focus-measure using a stepper motor connected to a microcontroller. The autofocusing module has been valuable for image recording over very long time periods (over 7-10 hours). Without autofocusing, manual adjustment of focus is required every 20-30 minutes, especially when recording microscale objects.

## EXPERIMENTAL RESULTS
### Cell Culture in Gel Microspheres

Our experimental process flow is shown in Figure 5. Initially, *E. coli* cells expressing green fluorescent protein (GFP) were mixed with the liquid agarose gel and dispensed on the droplet microfluidic system. The entire system was placed inside of environmental chamber. We also designed a PlexiGlass climate control environmental chamber to maintain a temperature between 36–37°C. After immobilizing the cell-laden gel microspheres at fixed locations, discrete drug droplets (with pre-defined concentrations of growth medium and drug solution) were mixed with specific cell-laden gel microspheres. The droplets were further diluted in buffer to a desired concentration using the methods of merging and mixing. The diluted droplets were brought to the specific agarose pads and allowed to be absorbed. Droplets of growth media were transported to the gel microspheres every 20 minutes to replenish the media lost by slow evaporation.

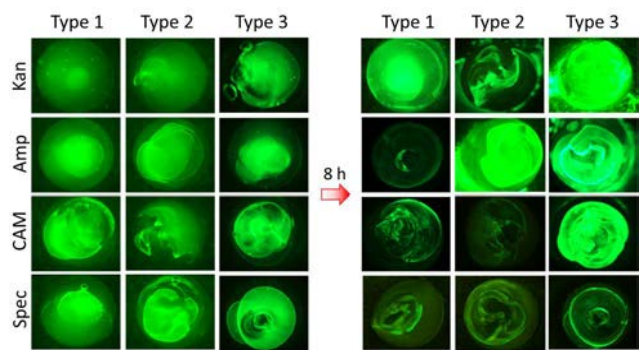

### Drug Testing

There are multiple classes of antibiotics that include inhibitors of cell-wall synthesis (β-lactams and glycopeptides), protein synthesis (aminoglycosides), and DNA synthesis (quinolones) [1]. Many derivatives of these drugs are used clinically to treat bacterial infections and include ampicillin, amoxicillin, tetracycline, penicillin, vancomycin, telavancin, ciprofloxacin, and tobramycin [2].

Our initial tests were performed with ampicillin. We remotely provided different concentrations of ampicillin solution to the solidified gel microspheres, and monitored the cell growth every 20 minutes under a microscope for the duration of the experiment. The proliferation or inhibition of cell growth was observed from the turbidity and measured optical density of gel microsphere for the next 6-10 hours. The optical density of each microsphere allowed us to measure the growth rate of bacteria. Microspheres exposed to a higher concentration of ampicillin solution displayed a lower fluorescent intensity.

Next, three different bacterial strains with resistance to different antibiotics were encapsulated within gel microspheres (Figure 6). Four different antibiotic solutions were supplied to each type of bacteria over a time period of 8 hours. By observing the spatiotemporal light intensities of gel microspheres, we were able to identify resistance types within 3 to 6 hours of culture. Further tests for bacterial resilience are being conducted by the sequential application of chemical stressors for finite time periods to understand when and how resilience emerges.

### Scanner-based Imaging of Persistent Bacteria

We developed a scanner-based imaging system to monitor the emergence of persisters, a specific type of resilient cells (Figure 2), within bacterial colonies cultured in plate assays. For this, we used a high-resolution flatbed scanner (Epson Perfection v750 Pro) connected to a computer. We imaged bacterial colonies grown in up to six 3-inch agarose plates (Figure 7). The recorded images were processed by a MatLab image processing program to accurately quantify and predict turbidity and cell growth/inhibition under different experimental conditions.

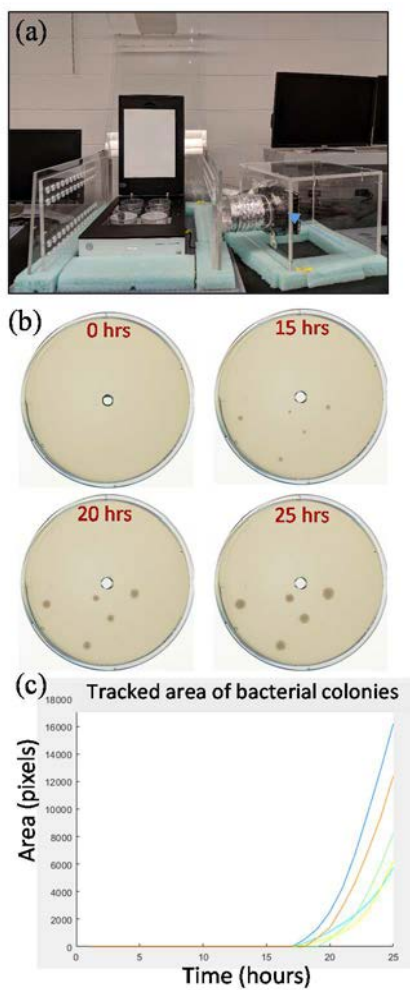

*Figure 7: (a) Image of our scanner-based plate assay setup within an environmental chamber to study bacterial persistence. (b) Our plate assays show the latency and eventual growth of persisters. (c) Our software program regularly scans up to six agar plates and extracts growth parameters for all the colonies within each plate.*

## Technology Limitations

There are limitations of the presented technology. For example, variability and heterogeneity in cell growth and survivability, even within the same population, are consistently observed in bacterial growth assays. Because of natural diversity, not all the cells within a culture display the same phenotype, including levels of drug resistance. Moreover, there are many scenarios where the data have poor quality or space-time resolution. Testing the search algorithm is unrealistic for large sets of experimental conditions [7-11]. Lastly, the results from *in vitro* experiments often do not replicate during *in vivo* studies.

## ACKNOWLEDGEMENTS

This work was supported by the U.S. Defense Threat Reduction Agency.

## REFERENCES

[1] S. M. Schrader, J. Vaubourgeix, C. Nathan, "Biology of antimicrobial resistance and approaches to combat it", *Science Translational Medicine*, vol. 12, issue 549, eaaz6992, 2020.

[2] M. S. Butler, D. L. Paterson, "Antibiotics in the clinical pipeline in October 2019", *The Journal of Antibiotics*, vol. 73, pp. 329-364, 2020.

[3] E. Palma, B. Tilocca, P. Roncada, "Antimicrobial Resistance in Veterinary Medicine: An Overview", *International Journal of Molecular Sciences*, vol. 21, no. 6, pp. 1914, 2020.

[4] A. Vasala, V. Hytönen, O. Laitinen, "Modern Tools for Rapid Diagnostics of Antimicrobial Resistance", *Frontiers in Cellular and Infection Microbiology*, vol. 10, pp. 308, 2020.

[5] S. Hassan, X. Zhang, "Microfluidics as an Emerging Platform for Tackling Antimicrobial Resistance (AMR): A Review", *Current Analytical Chemistry*, vol. 16, no. 1, pp. 41-51, 2020.

[6] E. M. Payne, D. A. Holland-Moritz, S. Sun, R. T. Kennedy, "High-throughput screening by droplet microfluidics: perspective into key challenges and future prospects", *Lab on a Chip*, vol. 20, pp. 2247-2262, 2020.

[7] T. Kong, N. Backes, U. Kalwa, C. Legner, G. J. Phillips, S. Pandey, "Adhesive Tape Microfluidics with an Autofocusing Module That Incorporates CRISPR Interference: Applications to Long-Term Bacterial Antibiotic Studies", *ACS Sensors*, vol. 4, no. 10, pp. 2638-2645, 2019.

[8] N. Jackson, L. Czaplewski and L. Piddock, "Discovery and development of new antibacterial drugs: learning from experience?", *Journal of Antimicrobial Chemotherapy*, vol. 73, pp. 1452-1459, 2018.

[9] X. Ding, Z. Njus, T. Kong, W. Su, C-H. Ho, S. Pandey, "Effective drug combination for *Caenorhabditis elegans* nematodes discovered by output-driven feedback system control technique", *Science Advances*, vol. 3, eaao1254, 2017.

[10] U. Ndagi, A. A. Falaki, M. Abdullahi, M. M. Lawal, M. E. Soliman, "Antibiotic resistance: bioinformatics-based understanding as a functional strategy for drug design", *RSC Advances*, vol. 10, pp. 18451-68, 2020.

[11] S. K. Gupta, B. R. Padmanabhan, S. M. Diene, R. Lopez-Rojas, M. Kempf, L. Landraud and J. M. Rolain, "ARG-annot, a new bioinformatic tool to discover antibiotic resistance genes in bacterial genomes", *Antimicrobial Agents and Chemotherapy*, vol. 58, pp. 212–220, 2014.